\documentclass[12pt,floatfix,showpacs,preprintnumbers,amssymb]{iopart}

\usepackage{graphicx}
\usepackage{pifont}

\graphicspath{{fig/}}
\usepackage{dcolumn}
\usepackage{bm}
\usepackage{float}
\usepackage{amssymb}
\usepackage{color}

\usepackage{textcomp}
\newcommand{\configAvert}{\bigl|\,%
\raisebox{-1.0ex}{\includegraphics[scale=0.3]{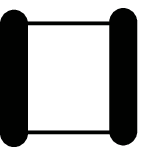}}\,%
\bigr\rangle}
\newcommand{\configAhor}{\bigl|\,%
\raisebox{-1.0ex}{\includegraphics[scale=0.3]{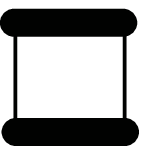}}\,%
\bigr\rangle}
\newcommand{\configB}{\Bigl|\,%
\raisebox{-2.0ex}{\includegraphics[scale=0.3]{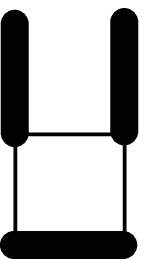}}\,%
\Bigr\rangle}

\newcommand{\configD}{\Bigl|\,%
\raisebox{-2.0ex}{\includegraphics[scale=0.3]{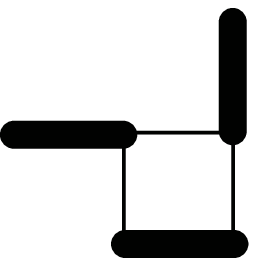}}\,%
\Bigr\rangle}

\newcommand{\configC}{\Bigl|\,%
\raisebox{-3.0ex}{\includegraphics[scale=0.3]{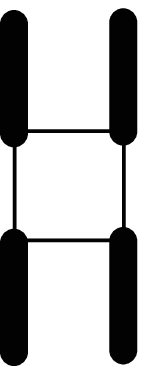}}\,%
\Bigr\rangle}
\newcommand{\configE}{\Bigl|\,%
\raisebox{-1.0ex}{\includegraphics[scale=0.3]{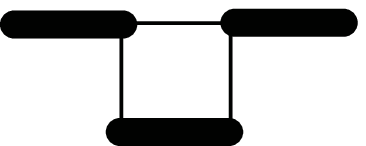}}\,%
\Bigr\rangle}

\newcommand{\braAhor}{\bigl\langle \,%
\raisebox{-1.0ex}{\includegraphics[scale=0.3]{VB2.eps}}\,%
\bigr|}
\newcommand{\braB}{\Bigl\langle \,%
\raisebox{-2.0ex}{\includegraphics[scale=0.3]{VB3.eps}}\,%
\Bigr|}
\newcommand{\braBB}{\Bigl\langle \,%
\raisebox{-2.0ex}{\includegraphics[scale=0.3]{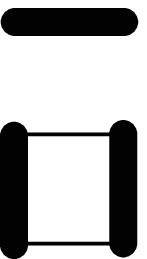}}\,%
\Bigr|}
\newcommand{\braD}{\Bigl\langle \,%
\raisebox{-2.0ex}{\includegraphics[scale=0.3]{VB5.eps}}\,%
\Bigr|}

\newcommand{\braC}{\Bigl\langle \,%
\raisebox{-3.0ex}{\includegraphics[scale=0.3]{VB7.eps}}\,%
\Bigr|}
\newcommand{\braE}{\Bigl\langle \,%
\raisebox{-1.0ex}{\includegraphics[scale=0.3]{VB8.eps}}\,%
\Bigr|}
\newcommand{\braCC}{\Bigl\langle \,%
\raisebox{-3.0ex}{\includegraphics[scale=0.3]{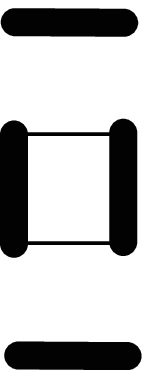}}\,%
\Bigr|}
\newcommand{\totsuka}[1]{\textcolor[rgb]{0,0,0}{#1}}
\newcommand{\nakata}[1]{\textcolor[rgb]{0,0,0}{#1}}

\begin{document}

\title[Extended Quantum Dimer Model ]
{Extended Quantum Dimer Model 
and novel valence-bond phases}

\author{Kouki Nakata and Keisuke Totsuka}

\address{Yukawa Institute for Theoretical Physics, 
Kyoto University, Kitashirakawa Oiwake-Cho, Kyoto 606-8502, Japan}
\ead{nakata@yukawa.kyoto-u.ac.jp}
\begin{abstract}
We extend the quantum dimer model (QDM) introduced by Rokhsar and Kivelson 
so as to construct a concrete example of the model 
which exhibits the first-order phase transition between different 
valence-bond solids
suggested recently by Batista and Trugman  
and look for the possibility of other exotic dimer states. 
We show that our model contains three exotic 
valence-bond phases ({\em herringbone}, {\em checkerboard} 
and {\em dimer smectic}) in the ground-state phase diagram and 
that it realizes the phase transition 
from the staggered valence-bond solid 
to the herringbone one.   
The checkerboard phase has four-fold rotational symmetry, 
while the dimer smectic, in the absence of quantum fluctuations, 
has massive degeneracy originating from partial ordering only 
in one of the two spatial directions. 
A resonance process involving three dimers resolves this massive degeneracy and
dimer smectic gets ordered (order from disorder).  
\end{abstract}
\noindent{\it Quantum Dimer Model,Valence-Bond Solid,\/}:
\submitto{J.Stat.Mech}

\maketitle

\section{Introduction}
\label{sec:intro}
Since P.W.~Anderson's paper\cite{anderson,andersonb,andersonc} 
in 1973, the quest for the resonating valence bond (RVB) 
state\cite{Hal,Takano,read,kohmoto} and exotic valence bond 
solids (VBS)\cite{ksr,rmos,huse,chan,ms} have been one of 
the recurrent themes in research on frustrated antiferro(AF) 
magnets\cite{lh,mg,ss}. 
One of the most studied problem of frustrated magnets would be 
the ground-state property of the $J_1$-$J_2$ 
Heisenberg model on a square lattice\cite{chandra,mel,sush,series} 
especially around the fully frustrated point $J_2/J_1=1/2$.
Chandra and Doucot\cite{chandra}, on the basis of a 1/S expansion 
for the sublattice magnetization, 
have predicted that the N\'{e}el order vanishes for $J_2/J_1$ $\simeq $ 0.38
and suggested that for a range of $J_2/J_1$ beyond 0.38
the ground state might be a spin liquid.  
Nishimori and Saika\cite{nishi} used the modified spin wave approximation
to conclude that a first-order phase transition 
from the N\'{e}el phase to the collinear AF phase\cite{Larkin} occurs.
Gelfand, Singh and Huse\cite{huse} have argued that system gets spontaneously
dimerized in the columnar pattern as has been predicted 
by Read and Sachdev\cite{RS}.
In spite of these intensive studies, nature of the ground state 
of the square-lattice $J_1$-$J_2$ Heisenberg model 
around the fully frustrated point $J_{2}/J_{1}=1/2$ is still 
controversial.   

One natural way to optimize the (short-range) antiferromagnetic correlation 
in quantum (i.e. low-spin) magnets is pairing spins at short distance 
into spin-singlet dimers. Therefore, it is tempting to consider the quantum 
dynamics within the spin-singlet subspace made up of 
all possible (short-range) dimer coverings.  
Quantum dimer model (QDM), 
\totsuka{%
which has been originally introduced 
by Rokhsar and Kivelson\cite{rk} to describe the low-energy physics 
of the square-lattice Heisenberg antiferromagnet, is now considered 
to capture a certain aspect of the low-energy dynamics of 
(frustrated) non-magnetic Mott insulators. 
}
The standard QDMs \'{a} la Rokhsar and Kivelson are defined 
in the Hilbert space of the nearest-neighbor dimer coverings 
of the lattice and consists only of processes which involve 
a single dimer pair. 
The square-lattice QDM thus defined is known to exhibit crystalline orders 
of the valence bonds and confined spinons except at 
the Rokhsar-Kivelson (RK) point\cite{rk}, where
a short ranged RVB state is realized (see Refs.~\cite{mis2,flip} 
for readable reviews of QDM). 

\totsuka{%
Of course, the Rokhsar-Kivelson QDM is a minimal model that describes 
the dynamics of singlet dimers; for instance, when one derives QDM 
from a given microscopic model by the overlap expansion, 
various higher-order terms (including dimer moves on larger loops) 
are generated.  Along this line, Ralko et al.\cite{ralko,kagomeQDM} 
has considered an extension of the QDM and investigated the impact 
of higher-order processes included. 
Among other attempts at extending QDM, 
Papanikolaou et al.\cite{papa} have introduced, 
on the basis of an analogy to 
the Pokrovsky-Talapov model\cite{pok} of fluctuating domain walls 
in two-dimensional classical statistical 
mechanics\cite{abh,pbak,mef,abha}, 
a two-dimensional microscopic model of interacting quantum dimers,  
which, in principle, involves infinitely many arbitrary 
parameters and dimer patterns.
}

\totsuka{%
Recently, Batista and Trugman\cite{batista} have undertaken 
a different microscopic approach and considered 
}
an $S=1/2$ $J_1$-$J_2$ Heisenberg model
with an additional term (four-spin exchange interactions)
that makes the model quasi-exactly solvable 
at the fully frustrated point $J_2/J_1=1/2$ 
\totsuka{%
(see Ref.~\cite{ksr} for a similar approach in spirit).
}
They have argued that any states having at least one singlet dimer per 
plaquette are ground states. 
Although one can easily see that the staggered- and 
the herringbone VBS\cite{papa} (dubbed `zigzag dimer' in 
Ref.\cite{batista}) are possible candidates of the ground states 
of their Hamiltonian, 
it is not clear when one of these VBSs becomes the unique ground state 
or what interaction controls the phase transition
between them and
it is known that the simplest QDM does not exhibit such a first-order 
transition\cite{Leung-C-R-96,qmc,Ralko-P-M-08}. 
Motivated by this question, we shall extend QDM so as to realize  
the first-order phase transition suggested in Ref.\cite{batista} 
and seek for other exotic dimer states
and then map out (\totsuka{a part of}) the phase diagram.
This is the main purpose of this paper.

\totsuka{%
This paper is structured as follows.  
In section \ref{sec:eqdm}, we study a certain region of EQDM 
in the absence of quantum fluctuations 
and, as the result, find a new disordered phase dubbed \textit{dimer smectic}. 
The main motivation comes from a first-order 
transition between the staggered VBS and the herringbone VBS which has been predicted 
recently by Batista and Trugman\cite{batista} 
to be controlled by a certain unspecified parameter `$\textit{g}$'. 
We show that by changing one of the coupling constants in our EQDM 
we can indeed drive the first-order transition. 
Within the usual QDM, both VBSs (staggered and 
herringbone) are exactly degenerate zero-energy states of 
the (classical) Hamiltonian and 
the single-plaquette resonance does not lift the degeneracy. 
}
\totsuka{%
In section \ref{sec:quantum-fluc}, we consider the effects of quantum
fluctuations. Specifically, 
we calculate the quantum correction to the ground-state energy  
for the three valence-bond states 
(herringbone, checkerboard VBSs and dimer smectic) 
which are degenerate in the absence of fluctuations (resonances) 
and see whether the degeneracy is resolved or not. 
We show that dimer smectic eventually gets ordered 
by a new resonance term which is equivalent to 
two successive actions of the familiar parallel dimer resonance.
}
In \ref{sec:definition}, we discuss an interesting mapping between 
the EQDM and a spin-1 model which is helpful in 
writing down the EQDM Hamiltonian in terms of a relatively small number of 
coupling constants.
In \ref{sec:cal},
we present the second-order calculation of the energy shift 
(eq.\ref{eqn:smectic}) caused by the $t'$ resonance effect for the dimer smectic.

\section{Extended Quantum Dimer Model}
\label{sec:eqdm}
Recently, Batista and Trugman\cite{batista} have introduced 
the following generalized $S=1/2$ $J_1$-$J_2$ Heisenberg model 
on a square lattice including an additional term 
(a four-spin exchange interaction) 
that makes a quasi-exact solution feasible at the fully 
frustrated point $J_2/J_1=1/2$: 
\begin{eqnarray}
{\cal H}_{\rm{BT}} &= {\cal{H}}_{J_1\rm{-}J_2} 
+ {\cal{H}}_{\rm{exchange}} \nonumber\\
&=J_1\sum_{\langle i,j\rangle} \vec S_{i} \cdot  \vec S_{j} 
+ \frac{J_1}{2}\sum_{\langle\!\langle i,j\rangle\!\rangle} 
\vec S_{i} \cdot  \vec S_{j}  \nonumber + \frac{J_1}{8}\sum_{\alpha}\left( 
P^{\alpha }_{ij}P^{\alpha }_{kl} 
+P^{\alpha }_{jk}P^{\alpha }_{il}+ P^{\alpha }_{ik}P^{\alpha }_{jl}
\right)   \nonumber\\
&= \frac{3J_1}{2}\sum_{\alpha } \cal{P}^{\alpha}  \;,
\end{eqnarray}
where $\langle i,j\rangle$ and 
$\langle\!\langle i,j\rangle\!\rangle$ denote the nearest neighbors 
and the second nearest neighbors, respectively, and 
$P_{ij}= S_{i}\cdot S_{j}$.
The index $\alpha $ labels the plaquettes and \textit{ijkl} are 
the four sites of each plaquette in the clockwise order. 
\totsuka{%
The four-spin exchange interaction ${\cal{H}}_{\rm{exchange}}$ is 
similar to the usual four-spin cyclic exchange\cite{ring} 
except that the sign of the last term is different. 
Since the operator $\cal{P}^{\alpha} $ projects the spin state 
of the plaquette $\alpha $ 
onto the subspace with total spin $S^{\alpha}_{T} =2$, 
}
it is clear\cite{batista} 
that any states having at least one singlet dimer on each 
plaquette can be the ground states. 
Of course, there are many other configurations\cite{batista}
where we have dimers on diagonal bonds or unpaired $S=1/2s$. 
\totsuka{%
Nevertheless, we will not consider these configurations hereafter. 
}

Though both the staggered- and the herringbone VBS (which is called  
`zigzag dimer' in Ref.\cite{batista})  
satisfy the condition for the ground states 
\totsuka{%
described above,
} it is not clear
when and how one of these VBSs is chosen as the unique ground state. 
Also the control parameter that drives the first-order 
\totsuka{%
transition from the 
staggered dimer to herringbone predicted in Ref.~\cite{batista} 
}
has not been identified 
yet either. 

\totsuka{
Motivated by these, we generalize the usual QDM and 
consider the following extended quantum dimer model (EQDM)}\footnote{%
\totsuka{%
The steps of defining our EQDM are outlined in Appendix.}}:
\begin{eqnarray}
 {\cal H}_{\rm{EQDM}}= \sum_{\Box}
\Biggl\{ 
&-t\, \configAvert \braAhor -t^{\prime}\,\configB\braBB -t^{\prime \prime}\,\configC\braCC  \nonumber \\
& +(-4J+2J^{\prime}+\lambda ) \,\configAhor\braAhor  -J \,\configD\braD  \nonumber\\
&-J^{\prime} \,\configB\braB  +(-2J+J'+\kappa ) \,\configE\braE \nonumber  \\
&+ J^{\prime} \,\configC\braC + (\rm{h.c.}) + (\rm{rotations}) 
+ (\rm{reflections})\Biggr\}.\
\label{eqn:EQDM}
\end{eqnarray}
\noindent
Except for the last term (with the coefficient $J^{\prime}$),
the above EQDM contains all possible dimer patterns having at least
one singlet dimer on each plaquette
(with the coefficients given by $-J$, $-J^{\prime}$, and
$(-2J+J'+\kappa )$); the last one appears in the process of deriving
the EQDM from the pseudo-spin ($S=1$) Hamiltonian described
in the Appendix. However, if we choose the coupling constants
as in (A.12)-(A.15), this configuration becomes higher in energy
and we may expect that
the EQDM hopefully becomes an effective Hamiltonian
for the generalized $S=1/2$ $J_1$-$J_2$ Heisenberg model
of Batista and Trugman\cite{batista}.

\nakata{We will study the region $J'=2J (>0)$, $0\leq \lambda < 2J$ and $\kappa <0$}
\totsuka{%
where the two valence-bond states -- the staggered dimer and 
the herringbone-- compete with each other in the ground state 
}
and see that the three-spin interaction $`\kappa '$  
in the pseudo-spin Hamiltonian (see Appendix for the definition) 
plays a role of the unspecified control parameter $`g'$ 
\totsuka{in Ref.~\cite{batista}}.
On the $\lambda =0$ line, 
\totsuka{%
a more exotic phase dubbed \textit{dimer
smectic} will be found. 
}

\subsection{Phase diagram in $J'=2J(>0)$, $0\leq \lambda < 2J$ and $\kappa <0$}
\label{sec:diagram}
In this region ($J'=2J$ ($>$0)), the EQDM is rewritten as
\begin{eqnarray}
 {\cal H}_{\rm{EQDM}}&= \sum_{\Box}
\Biggl\{ 
-t\, \configAvert \braAhor -t^{\prime}\,\configB\braBB 
-t^{\prime \prime}\,\configC\braCC \nonumber\\
 &\ \ \ \ \ \ \ \ \ \ \ \ \ +\lambda  \,\configAhor\braAhor -J \,\configD\braD -2J \,\configB\braB \nonumber\\
  &\ \ \ \ \ \ \ \ \ \ \ \ \  +\kappa  \,\configE\braE 
+ 2J \,\configC\braC \nonumber\\
&\ \ \ \ \ \ \ \ \ \ \ \ \  + (\rm{h.c.}) + (\rm{rotations})+ (\rm{turn over})\Biggr\} .
\end{eqnarray}
We restrict our investigation to the region $\lambda \geq 0$ 
because otherwise the staggered dimer and the herringbone 
are not competing, and such situation is not relevant to our current purpose.

\begin{figure}[h]
\begin{center}
\includegraphics[width=12cm, clip]{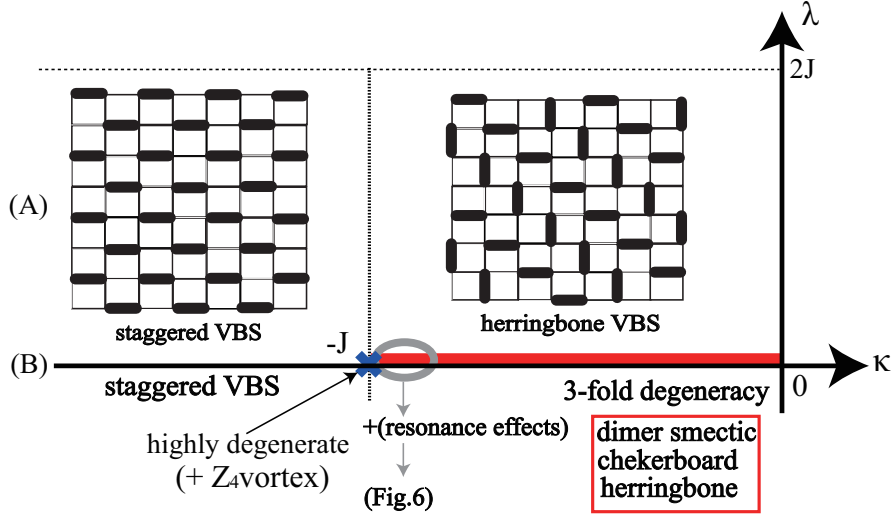}
\caption{(Color online) Phase diagram of EQDM 
for $t=t^{\prime}=t^{\prime \prime}=0$, $0\leq \lambda <2J$.  
Basically $\kappa $ controls the phase transition from
staggered-VBS to herringbone-VBS.  
On $\lambda =0,\kappa >-J$ line
checkerboard-VBS and dimer smectic emerge and are degenerate
with herringbone-VBS. \label{fig:classical-PD}}
\end{center}
\end{figure}

\begin{figure}
\begin{center}
\includegraphics[width=10cm, clip]{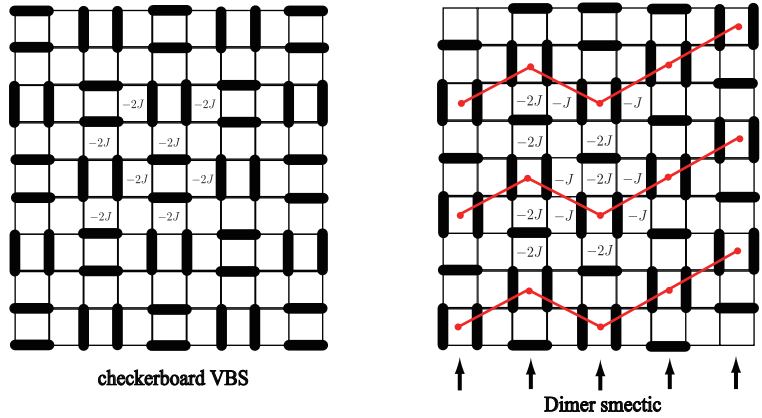}
\caption{(Color online) [Left] Checkerboard. This VBS is characterized by 
a checkerboard pattern of parallel dimers 
and has the four-fold rotational symmetry.
[Right] Dimer smectic. 
This is realized by arranging ({b}) plaquettes in Fig.\ref{fig:allowed-plaquette-states}
in such a way that the number of ({b}) plaquettes is maximal within each column
and then connecting these `ordered' columns horizontally
by the herringbone-like ({d}) plaquettes in Fig.\ref{fig:allowed-plaquette-states}.
\label{fig:stripe-B}}
\end{center}
\end{figure}

\totsuka{%
In the first step, we consider the case where quantum fluctuations 
(i.e. the resonances $t$, $t^{\prime}$ and $t^{\prime\prime}$) are absent. 
}
The phase diagram for $t=t^{\prime}=t^{\prime \prime}=0$ is given 
in Fig.\ref{fig:classical-PD}.  
\noindent%
(A) $0<\lambda<2J$; 
it is clear that the phase transition from the staggered-VBS to the herringbone-VBS 
occurs by changing the parameter $\kappa $, which is the coefficient
of three-spin interaction in the pseudo-spin Hamiltonian given in Appendix. 
The critical value is $\kappa_{\rm{c}}=-J $.
Namely, the three-spin interaction  `$\kappa $' in the pseudo-spin Hamiltonian 
plays a role of the unspecified control parameter `$g$' in Ref.\cite{batista}.
In this sense, EQDM may be an effective Hamiltonian for a
$J_1$-$J_2$ Heisenberg model with a four-spin interaction 
introduced by Batista and Trugman in this region. 
It is worth mentioning that the usual QDM cannot describe 
the transition from staggered to herringbone VBS, since both are 
exactly degenerate zero-energy eigenstates of the QDM Hamiltonian 
regardless of the parameters.
\noindent%
(B) $\lambda =0$; Two novel kinds of VBS, {\em the checkerboard VBS} 
(Fig.\ref{fig:stripe-B}) and {\em the dimer smectic} 
(Fig.\ref{fig:stripe-B}), emerge and are degenerate 
with the herringbone-VBS ($\kappa \geq  -J$). 
In the next section, 
we will discuss how to resolve this degeneracy by the 
effects of resonance ($t,t^{\prime}$,$t''$) in detail.

On the $\kappa =-J$ and $\lambda =0$ point the ground state is highly degenerate and Z$_4$-vortex (Fig.\ref{fig:Z4}) emerges.

\subsection{The Features of VBS Phases}
\label{sec:feature}
Now let us discuss the main features of these VBSs. 

\noindent%
(i) \textit{Herringbone};   
The hallmark of this state is that it contains 
no parallel dimer.  In fact, the herringbone shares 
this important property with the well-known staggered VBS\cite{rk}; 
there is no {\em flippable} (with respect to $t$, $t^{\prime}$ and $t^{\prime \prime}$) 
dimers in both VBSs.
In this sense, both the herringbone and the staggered VBS are 
total free from the resonance and robust. 

\noindent%
(ii) \textit{Checkerboard VBS}\cite{check2}; this VBS is made up of 
a checkerboard pattern of two kinds of parallel dimers 
and has the four-fold rotational symmetry . 

\begin{figure}[h]
\begin{center}
\includegraphics[width=7cm, clip ]{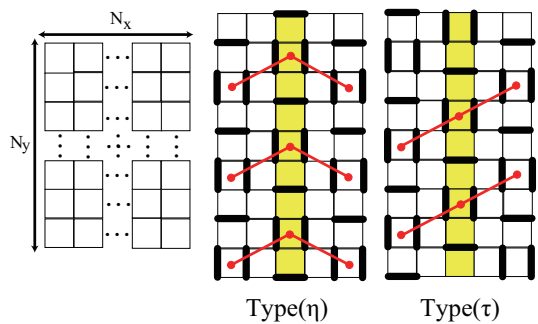}
\caption{(Color online) Two possible configurations of three 
neighboring columns in dimer smectic.  
These two patterns are the building blocks in
forming highly degenerate order in the horizontal direction. %
\label{fig:two-config}}  
\end{center}
\end{figure}

\begin{figure}
\begin{center}
\includegraphics[width=7cm, clip]{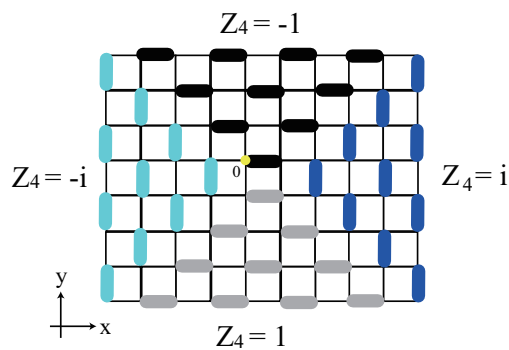}
\caption{(Color online) Illustration of the $Z_4$ vortex. 
Four dimer domains are characterized by four different values ($1,-1,i,-i$)
of order parameter ${\cal O}_{\mathbb{Z}_{4}} $.} %
\label{fig:Z4}
\end{center}
\end{figure}

\noindent%
(iii) \textit{Dimer smectic}; 
this VBS corresponds in a sense to the intermediate 
between the checkerboard and the herringbone, 
which means that this
is realized by arranging ({b}) plaquettes in Fig.\ref{fig:allowed-plaquette-states}
in such a way that the number of ({b}) plaquettes is maximal within each column
and then connecting these `ordered' columns horizontally
by the herringbone-like ({d}) plaquettes in Fig.\ref{fig:allowed-plaquette-states}.
The horizontal order is formed so as to break the translation
symmetry. In other words, there are two ways (type-($\eta $) and ($\tau
$) in Fig.\ref{fig:two-config}) of forming the herringbone-like 
order horizontally.
If the integers $N_{x}$ and $N_{y}$ are the linear dimensions of the lattice
(the lattice spacing is set to unity) and
we set $N_{\eta}$ and $N_{\tau}$ the numbers 
of the stripes of type-($\eta$) and ($\tau$),
they  must satisfy
\begin{equation}
 2N_\eta +2N_\tau =N_x. 
\label{eqn:constraint-Neta-Ntau}
\end{equation}
From this, one sees that 
the horizontal herringbone-like order type-($\eta$ ) and ($\tau$) 
not only break the translation symmetry but also 
lead to extensive GS degeneracy since any combinations of
($\textit{N}_{\eta },\textit{N}_{\tau }$) are permitted if they only
satisfy eq.(\ref{eqn:constraint-Neta-Ntau}). 
This huge degeneracy comes from the fact that pure (local) energetics 
cannot determine the ground state uniquely and is reminiscent of 
similar degeneracy in such geometrically frustrated magnets as 
the Kagom\'{e} antiferromagnet\cite{Chubukov} 
and the pyrochlore antiferromagnet\cite{Moessner-C-98}. 
\begin{figure}[h]
\begin{center}
\includegraphics[width=10cm,clip]{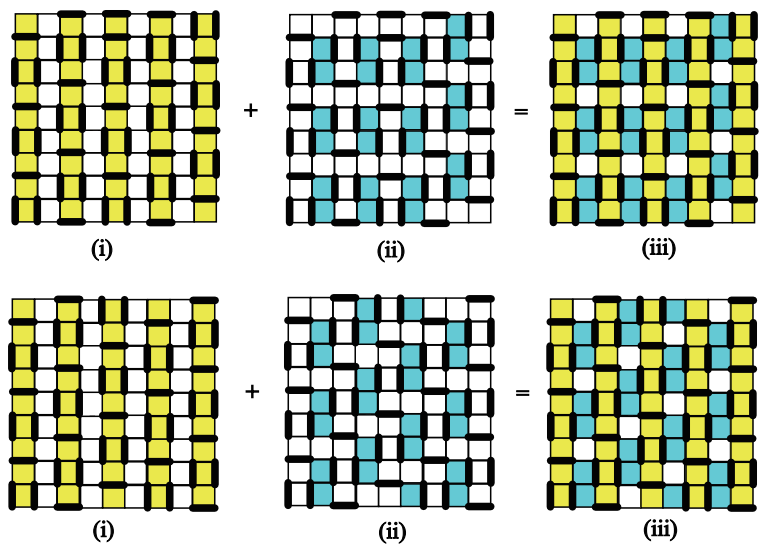}
\caption{(Color online) Plaquette tiling of stripe. 
Hatched columns in panel-(i) may be viewed as one-dimensional solids
consisting of type-(\textit{b}) plaquettes and these columns are 
connected to each other by `binding energy' provided by 
the interaction energy of herringbone-like (hatched) plaquettes in panel-(ii).%
\label{fig:plaquette-tiling}}
\end{center}
\end{figure}
\noindent%
(iv) \textit{Z$_4$-vortex}; this VBS becomes one of the ground states
only at the special point $\lambda =0,\kappa =-J$ (shown in Fig.\ref{fig:classical-PD} 
as `highly degenerate').  
This phase consists of four domains each of which assumes 
one of the four possible staggered VBS patterns (see Fig.\ref{fig:Z4}). 
We call this phase $\mathbb{Z}_4${\em -vortex}\footnote{%
This is different from what is discussed in the context of 
the deconfined criticality\cite{Levin-S-04} in that there is no 
unpaired spin-1/2 which is responsible for the stabilization 
of the N\'{e}el phase out of the columnar VBS.
}
because the following order parameter ${\cal O}_{\mathbb{Z}_{4}}$ 
in fact changes its value like a $\mathbb{Z}_{4}$-variable 
($1\rightarrow i \rightarrow -1 \rightarrow -i$) 
as we move around the origin (see Fig.\ref{fig:Z4}):
\begin{eqnarray}
{\cal O}_{\mathbb{Z}_{4}} 
\equiv & \frac{1}{N_{\rm{site}}}\sum_{\vec r} 
(-1)^{\vec r_x+\vec r_y}
\Bigl\{
( \vec S_{\vec r} \cdot  \vec S_{\vec r + \vec e_x}
- \vec S_{\vec r + \vec e_y } \cdot  \vec S_{\vec r + \vec e_x +\vec e_y})  \nonumber \\
&\ \ \ \ \ \ \ \ \ \ \ \ \ \ \ \ \ \ \ \ \ \ \ \ \ \   +i( \vec S_{\vec r} \cdot  \vec S_{\vec r + \vec e_y}- \vec S_{\vec r +
 \vec e_x } \cdot  \vec S_{\vec r + \vec e_x +\vec e_y}) 
\Bigr\} \; .
\end{eqnarray}

\section{Effects of Quantum Fluctuations}
\label{sec:quantum-fluc}
\subsection{Energy corrections from resonance terms}
\begin{figure}[h]
\begin{center}
\includegraphics[width=12cm, clip ]{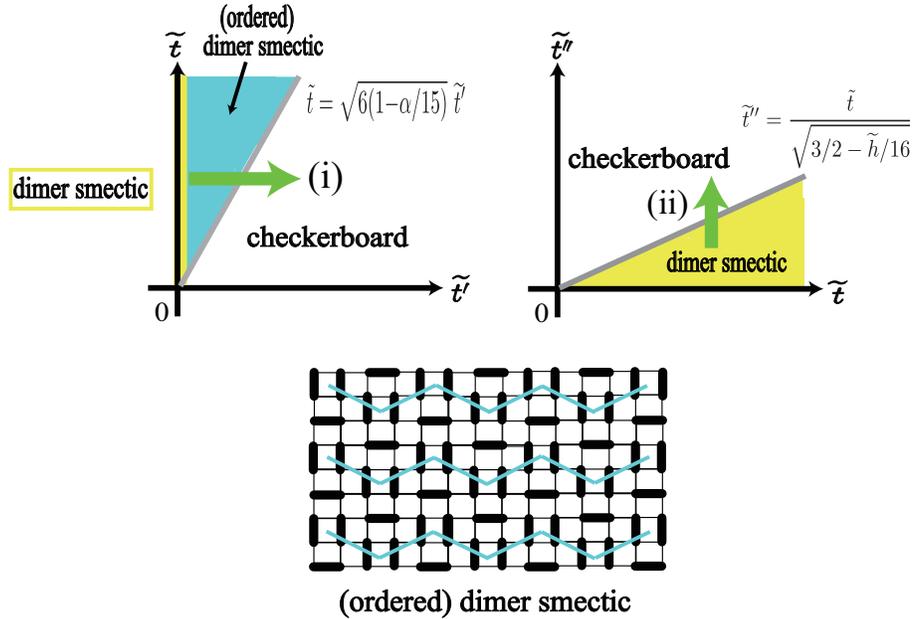}
\caption{(Color online) Schematic phase diagram of EQDM with ($\tilde t'$-$\tilde t$) plane and ($\tilde t$-$\tilde t''$)plane
for $J'=2J (>0),\lambda =0,\kappa =-(J+h)$. Parameter  $h$ is a small  negative value. 
Dimer smectic gets ordered only by resonance $\tilde t'$ (see arrow (i)).
 Because the phases on either side of the boundary have totally different symmetries,
we may expect that the transition is of first order.%
\label{fig:phase-diag-1}}
\end{center}
\end{figure}

In this section, we investigate how the degeneracy among 
the three valence-bond phases (herringbone, 
checkerboard and dimer smectic) found in the previous section 
is resolved for $J'=2J (>0),\lambda =0$ and $\kappa =-(J+h)$ 
(circled region in Fig.\ref{fig:classical-PD}) paying particular 
attention to the fate of the huge ground-state degeneracy in the dimer 
smectic. 
First let us recover the resonance terms $t$, $t^{\prime}$ and $t^{\prime \prime}$ 
as perturbation to the diagonal part:
\begin{eqnarray}
 V_{\rm{perturb}}&= \sum_{\Box}\left\{-t\, \configAvert \braAhor -t^{\prime}\, \configB \braBB
-t^{\prime\prime} \,\configC\braCC \right\}   \nonumber\\
&\ \ \ \ \ \ \ \ \ \ \ + (\rm{rotations}) + (\rm{h.c.}).
\end{eqnarray}
The resonance terms $t$, $t^{\prime}$ and $t^{\prime \prime}$ are assumed to be very 
small (i.e. $0< t,t^{\prime},t''\ll J$), and will be used as the small
parameters in perturbation theory. These perturbing terms come into play
for the first time at the second order and we have to consider 
the effects of resonance terms up to this order. 
\begin{eqnarray}
 E_{\rm{resonance}}^{\rm{check}}&=
-JN_xN_y - \frac{t'^2}{4J}\frac{N_xN_y}{2} 
-\frac{t^2}{12J}\frac{N_xN_y}{4}-\frac{t''^2}{6J}\frac{N_xN_y}{4}. \\ 
 E_{\rm{resonance}}^{\rm{smectic}}&=
-JN_xN_y - \frac{t'^2}{4J-2h}N_{\eta }\frac{N_y}{3} 
- \frac{t'^2}{10J}N_{\eta }\frac{N_y}{3}  
- \frac{t'^2}{7J-h}N_{\tau }\frac{2N_y}{3}  \nonumber \\  
&\ \ \ \ \ -\frac{t^2}{4J}\frac{N_x}{2}\frac{N_y}{3}-\frac{t''^2}{16J-2h}\frac{N_xN_y}{6}. 
\label{eqn:smectic} \\
 E_{\rm{resonance}}^{\rm{herring}}&=-JN_xN_y.
\label{eqn:herr}\
\end{eqnarray}
The pair of integers $(N_{\eta},N_{\tau})$ has been defined 
in section 3 and satisfies 
\begin{equation}
2N_{\eta }+2N_{\tau }=N_x \; .
\end{equation}
We now rescale the couplings as $\tilde{t}\equiv t/J$ and 
$\tilde{h}\equiv h/J$ and define $\bar {N_\eta}\equiv  N_\eta /N_x$ 
so that $\bar {N_\eta }$ must satisfy
$0 \leq \bar {N_\eta }\leq 1/2$.  
The parameter $|h|$ is assumed to be very small ($|h|\ll J$)
and we approximate $\tilde h^2 \sim 0$.  
Then, the scaled energy 
$\widetilde E_{\rm{resonance}}^{\rm{smectic}}\equiv
E_{\rm{resonance}}^{\rm{smectic}}/J$ is given by
\begin{equation}
\frac{\widetilde E_{\rm{resonance}}^{\rm{smectic}}}{N_xN_y}\simeq 
-1 - \frac{\tilde t^2}{24} 
- \frac{\tilde t'^2}{21}\left(1+\frac{\tilde h}{7} \right)-\frac{\tilde
t'^2}{140}\bar  N_\eta \left(3+\frac{55}{14}\tilde h\right)
-\frac{\tilde t''^2}{96}\left(1+\frac{\tilde h}{8} \right).  
\end{equation}
It is apparent that only the resonance $\tilde t'$ depends on the number of
stripe type (($ \eta $) and ($ \tau $)) in dimer smectic (fourth term in the right hand side).
Since $|h|$ is so small, we may safely assume that the coefficient of 
$\bar N_\eta $ is negative. 
Then, one can easily see that quantum fluctuations ($t^{\prime}$) form 
a minimum of the energy of the dimer smectic 
at $\bar N_\eta =1/2$ thereby select one particular configuration,  
which is shown in the lower panel of Fig.\ref{fig:phase-diag-1},  
as the ground state. 
This is because when $|h|$ is small enough,
type($\eta$) configuration can gain more resonance ($t^\prime$) energy 
than the type($\tau $) one and the total energy is minimized when  
all the columns are of the type($\eta$). 

Thus resonance $\tilde t'$ which is equivalent to two successive actions of the
usual parallel dimer resonance ($\tilde t$) and ignored 
in the standard QDM plays a special role; it resolves the high degeneracy through the order-by-disorder mechanism\cite{obdo-1,obdo-2}.  
 
\begin{figure}[h]
\begin{center}
\includegraphics[width=12cm, clip ]{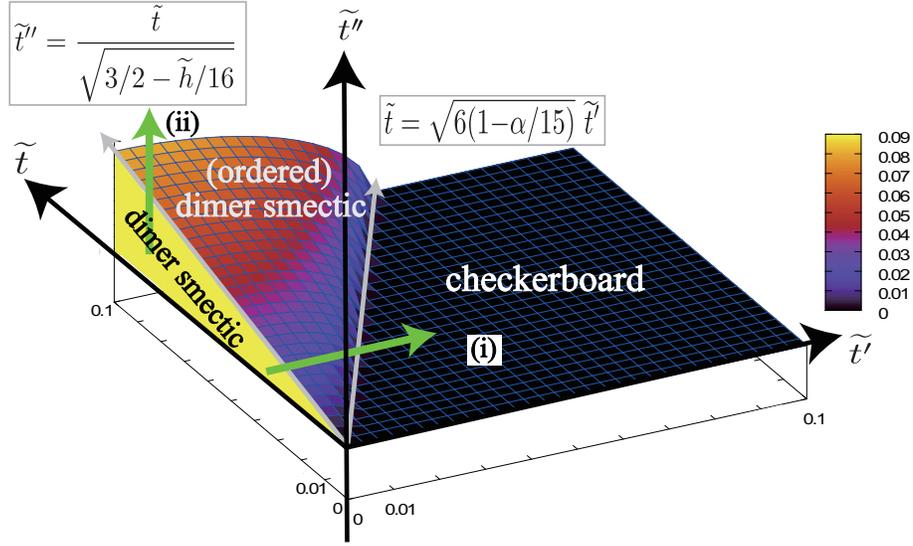}
\caption{(Color online) Schematic phase diagram of EQDM with $t$-$t'$-$t''$axis
for $J'=2J (>0),\lambda =0,\kappa =-(J+h)$. 
Parameter $h$ takes small negative values. Highly degenerate dimer smectic exists only within a region on the $\tilde{t'}=0$ plane 
specified by $\tilde{t}^{\prime\prime}< \tilde t ( \frac{3}{2} - \frac{\tilde h}{16}      )^{-1/2}$.  
In the region shown as `(ordered) dimer smectic', a unique ground state is selected by the resonance $\tilde{t}^{\prime}$.  
Boundary between ordered dimer smectic and checkerboard VBS is given by eq.(33). 
\label{fig:goodphase}}
\end{center}
\end{figure}

Then the energy of ordered dimer smectic is evaluated as
\begin{equation}
 \frac{\widetilde E_{\rm{resonance}}^{\rm{smectic}}}
{N_xN_y} \simeq 
-1 - \frac{\tilde t^2}{24} - \frac{\tilde t'^2}{120}\alpha 
-\frac{\tilde t''^2}{96}\left(1+\frac{\tilde h}{8} \right). 
\end{equation}
\begin{equation}
\alpha \equiv 7+ \frac{5}{2}\tilde h \; .
\end{equation}
By comparing the energies of the three phases, 
we can draw the phase diagram of EQDM in this region 
(Fig.\ref{fig:goodphase}).
\begin{eqnarray}
 \widetilde E_{\rm{resonance}}^{\rm{smectic}} 
< \widetilde E_{\rm{resonance}}^{\rm{check}} 
 &\Longleftrightarrow  
\tilde t^2  > 
6\left( 1- \frac{\alpha }{15} \right)\tilde t'^2
+\left( \frac{3}{2}-\frac{\tilde h}{16} \right)\tilde t''^2.
\label{eqn:region}
\end{eqnarray}
As shown in Fig.\ref{fig:phase-diag-1} dimer smectic gets ordered by resonance $t'$ on ($\tilde t'$-$\tilde t$) plane 
and resonance $\tilde t'$ drives the phase transition from dimer smectic to checkerboard VBS through ordered dimer smectic.
On the other hand, on ($\tilde t$-$\tilde t''$) plane (without $t'$ resonance) 
because there is no resonance $t'$ effects that dimer smectic does not get ordered and
the phase transition from dimer smectic to checkerboard VBS occurs 
without  ordered dimer smectic phase. 

Because the phases on either side of the boundary have totally different symmetries,
we may expect that the transition is of first order.

\subsection{Finite temperatures}

It would be interesting to consider how the above 
one-dimensional solids in dimer smectic phase  melt into disordered states 
at finite temperatures. 
In fact, this kind of columnar structures reminds 
us of the smectic metal state in strongly-correlated 
systems\cite{stripe1,stripe2} and the sliding phases 
in liquid crystals\cite{OHern-sliding} or in frustrated spin 
systems\cite{Starykh,Sindzingre}. Below, we give a brief 
discussion about the possibility of the sliding behavior 
in the dimer smectic phase.

\begin{figure}[h]
\begin{center}
\includegraphics[width=7cm,clip]{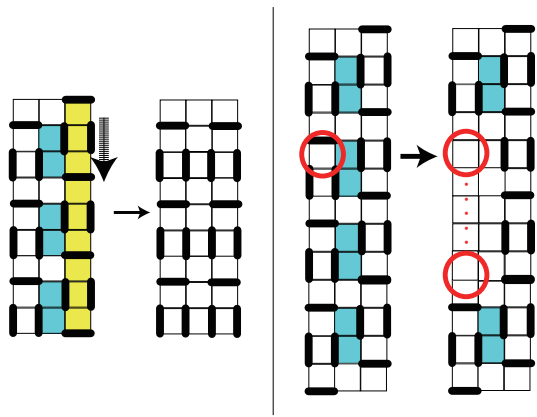}
\caption{(Color online) 
[Left] Shift of a solid column (the rightmost one) in the vertical direction 
destroys the herringbone-like order (hatched plaquettes 
in the middle column). 
The necessary energy cost is estimated as 
 $\Delta  E_{\rm {shift}}^ {\rm{smectic}} \sim O(N_y)$.
 [Right] Destruction of a columnar (1D) solid {\em itself} 
occurs by the insertion of wrong  plaquettes 
into the column. 
The energy to destroy one-dimensional solid itself 
is estimated as  $\Delta  E_{\rm {column}}^ {\rm{smectic}} \sim O(1)$. 
 \label{fig:E-slide-1}}
\end{center}
\end{figure}

The energy cost necessary to slide 
a column (a one-dimensional solid) vertically and destroy 
the dimer-smectic (i.e. herringbone-like) order is estimated as 
(see Fig.\ref{fig:E-slide-1})
\begin{equation}
\Delta E_{\rm{shift}}^{\rm{smectic}}\sim O(N_y) \; .
\end{equation}  
On the other hand, 
the energy to destroy a one-dimensional solid itself is estimated as 
\begin{equation}
\Delta E_{\rm{column}}^{\rm{smectic}}\sim O(1),
\end{equation} 
because we can adjust dimer configurations so as to minimize the effects 
of breakdown in one-dimensional solid (see Fig.\ref{fig:E-slide-1}).
These values suggest that as the temperature is increased, 
one-dimensional solids themselves are destroyed in the first place 
and exclude the possibility of the phase of fluctuating one-dimensional
solids. 

\section{Summary }
\label{sum}
In the usual QDM of Rokhsar and Kivelson , the staggered- and the herringbone VBS 
should be degenerate, because both VBSs are zero-energy states and have no plaquettes on which 
resonance term ($t$) and diagonal term (pair dimer) can act.
 
In the present paper, we have extended QDM so as to describe the phase transition
from the staggered VBS to  the herringbone VBS.
We can include all (nearest-neighbor) dimer configurations which have at least one dimer in each plaquette.
Therefore, EQDM is considered in a sense to be a generalization of $S=1/2$ the $J_1$-$J_2$ model
with four-spin exchange interactions\cite{batista} at its fully frustrated point ($J_1=2J_2$).
 
We then mapped out the phase diagram of EQDM in the region
where both the staggered VBS and the herringbone VBS exist as the ground states
($J'=2J(>0),\lambda =0,\kappa =-(J+h)$)
by the second-order perturbation theory 
in quantum fluctuations $t$, $t^{\prime}$ and $t^{\prime \prime}$.
We have found that the three-spin interaction $`\kappa^{\prime}$ 
in the pseudo-spin Hamiltonian plays a role of 
the unspecified control parameter `$\textit{g}$' suggested 
by Batista and Trugman 
which drives the first-order phase transition between the above two VBS 
phases. 
We have also found that there does exist  a new VBS phase called the dimer smectic. 
This novel phase forms period-3 structure only in one (say, vertical)
direction, and each ordered columns are connected 
by the herringbone-like plaquettes.
The order in the horizontal direction
is formed so as to break the translational symmetry
and leads to huge ground-state degeneracy. 
A resonance process involving three dimers ($t'$)  resolves this massive degeneracy and
dimer smectic eventually gets ordered 
(see the alternating pattern in Fig.\ref{fig:phase-diag-1}) 
through the order-by-disorder mechanism.  

Last, in this paper, we did not pursued the (approximate) realization 
of our EQDM in specific spin models. 
Though our original motivation of introducing EQDM is to present 
a specific model which realizes a first-order
staggered-herringbone transition predicted in Ref.\cite{batista} 
and exhibits a novel dimer smectic phase, 
this phase unfortunately does not satisfy the 
condition for the ground state of the generalized $J_1$-$J_2$ spin model 
of Batista and Trugman 
(some plaquettes in the dimer smectic do not contain dimer bonds).  
Therefore it is interesting to look for spin Hamiltonians which exhibit 
the dimer smectic.

\ack
We thank K.~Harada for useful comments at the early stage
of this study and 
K.~Penc for discussions on the Batista-Trugman model.
The author (K.T.) was supported by Grant-in-Aid
for Scientific Research (C) 20540375 and that on
Priority Areas ``Novel States of Matter Induced by Frustration''
(No.19052003) from MEXT, Japan.

\appendix
\section{Definition of Extended QDM}
\label{sec:definition}
\totsuka{%
In this appendix, we describe an interesting mapping 
between the EQDM (\ref{eqn:EQDM}) and a (pseudo)spin-1 Hamiltonian with 
multi-spin interactions. 
This mapping helps us reduce the number of free parameters 
and keep a small number of relevant dimer configurations. 
in the spirit close to that of 
Moessner, Sondhi and Fradkin(see Ref.\cite{map}).}

\begin{figure}[h]
\begin{center}
\includegraphics[width=10cm, clip ]{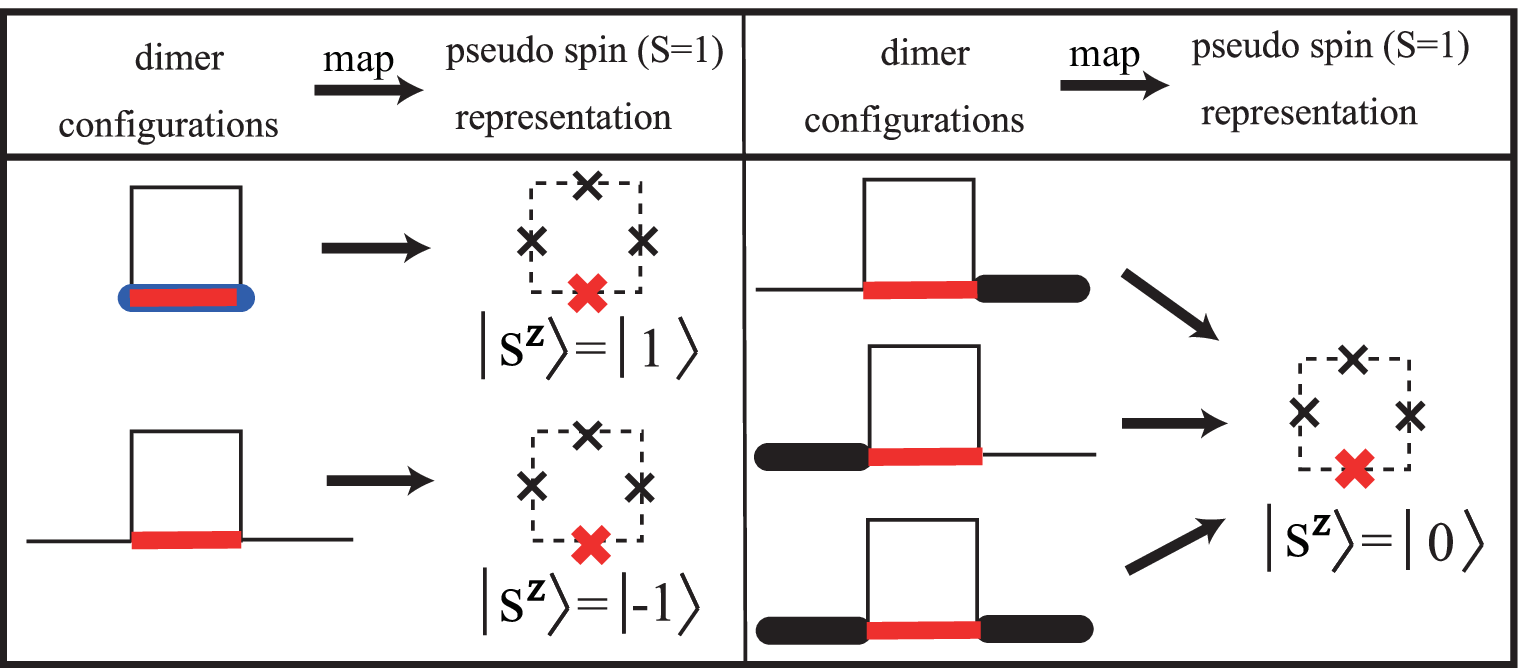}
\caption{(Color online) Definition of the eigenstates $|$ $\pm$
 1 \textrangle \, $|$0 \textrangle \ .  $|$1 \textrangle \ is located on the center
 of link which is occupied by a dimer, and  $|$0 \textrangle ,
 $|$-1\textrangle \ are located on the center of links where dimers are
 absent. $|$-1\textrangle \ is placed on the link where either sides links
 are not occupied by dimer, and  $|$0 \textrangle \  is the others. %
\label{fig:Def-S1-states}}
\end{center}
\end{figure}

The key step in establishing the mapping is to assign 
the eigenstates $|1\rangle$, $|0\rangle$ and $|-1\rangle$ of 
the pseudo-spin-1 operator $s^{z}$ to the dimer configurations 
{\em not} on a single bond 
but on an extended cluster containing the bond on which the spin-1 
is defined (see Fig.\ref{fig:Def-S1-states}).  
The state $|1\rangle$ is assigned when the bond is occupied by
a dimer and either $|0\rangle$ or $|-1\rangle$ is assigned 
otherwise. 
This rule connects the dimer configurations to those of the 
$S=1$ pseudo-spins. 

In principle, one could have taken another strategy. 
Namely, one could have used the $S=1/2$ 
operators to map a QDM containing large resonance loops and long-range 
(dimer-dimer) interactions\cite{papa} onto an $S=1/2$ (pseudo)spin Hamiltonian 
with multi-spin- (six-spin and more) and long-range interactions.  
However, this may lead to the complexity of the interactions and 
hinder further analyses of the resulting (pseudo-spin) Hamiltonian 
as well as loss of local Ising gauge invariance. 
If we use the spin-1 mapping, on the other hand, we can still 
work within the space of Hamiltonians with only \textit{short-range interactions}
in the sense that the spin-spin interactions exist only among four spins
forming a plaquette. 
In this sense, our EQDM may be thought of as  
a minimal model of generalized QDMs.  

\begin{figure}[h]
\begin{center}
\includegraphics[width=12cm, clip ]{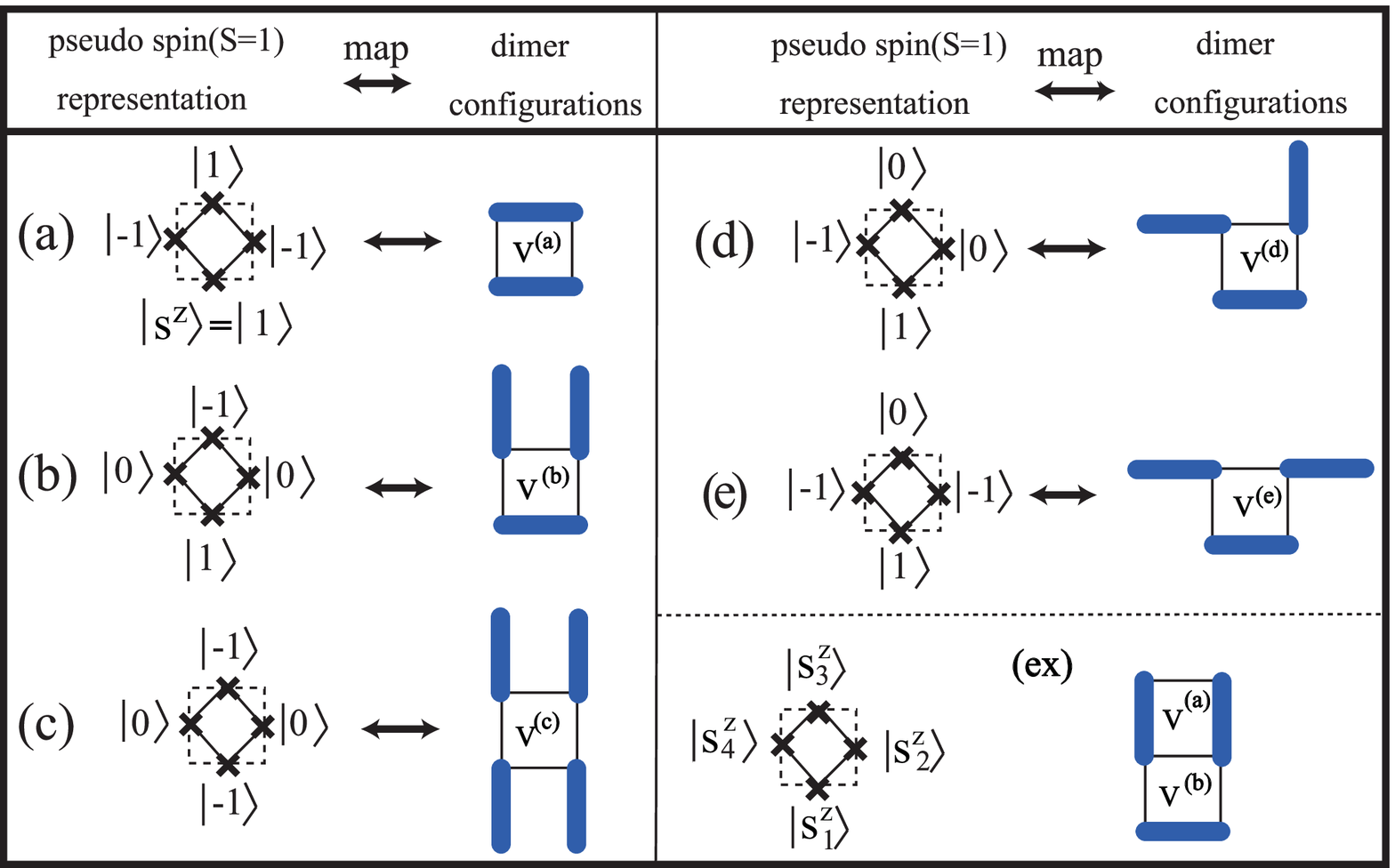}
\caption{(Color online) 
\totsuka{%
Five states allowed by the hardcore
 constraint and $V_{\rm EQDM}$. 
Some of the configurations allowed by hardcore constraint have been dropped 
for the reason that they have zero coefficient in the diagonal part 
of the Hamiltonian. The pseudo-spins are located on the middle of each
 link. 
}
\label{fig:allowed-plaquette-states}}
\end{center}
\end{figure}

Obviously, the naive state space spanned by the pseudo-spin 1s are much larger 
than that of the QDM and we have to implement the hardcore dimer 
constraint in the pseudo-spin language:
\begin{equation}
\sum_{+}     \Theta  (s_{z})=1 \quad
\mbox{ with }
\;
 \Theta (x)\equiv \cases{1&\rm{when $0<x$}\\
0&\rm{when $x\leq 0$}\\} \; .
\label{eqn:spin-cinstraint}
\end{equation}
When the hardcore constraint is neglected, a single plaquette can take 
$3^{4}$ different configurations. 
When the hardcore dimer constraint that at each site only one of the 
four links emanating from it can be occupied by a dimer
and coefficient of diagonal terms ($V_{\rm{EQDM}}$ discussed in the next chapter) into account, 
then relevant dimer arrangements of a plaquette are restricted to 
the five patterns shown in Fig. \ref{fig:allowed-plaquette-states}. 
At this point, one may notice that some dimer configurations 
allowed by the hardcore dimer constraint are missing in the patterns 
shown in Fig.~\ref{fig:allowed-plaquette-states}.
For instance, a spin-1 state 
$|s^{z}_1\rangle|s^{z}_2\rangle|s^{z}_3\rangle|s^{z}_4\rangle$ 
with $s^{z}_{1}{=}s^{z}_{2}{=}s^{z}_{3}{=}s^{z}_{4}{=}0$ 
(a pinwheel-like configuration) is permissible 
but the coefficient of the diagonal term corresponding to this state 
is zero as will be seen in  eq.(\ref{eqn:Veqdm}). 
A similar argument enables us to drop other constraint-allowed 
configurations from Fig.~\ref{fig:allowed-plaquette-states}. 

Having defined the three $S=1$ pseudo-spin eigenstates 
in terms of the dimer configurations on an extended (three-bond)
cluster, 
we are at the position of defining the diagonal part of our 
pseudo-spin Hamiltonian:
\begin{equation} 
V_{\rm{EQDM}}=J s_i^z s_{i+1}^z +J's_i^z s_{i+2}^z  
 +\kappa s_i^z s_{i+1}^z s_{i+2}^z  +\lambda s_i^z s_{i+1}^z s_{i+2}^z s_{i+3}^z \; ,  
\label{eqn:Veqdm}
\end{equation}
where the first two are the well-known $J_{1}$-$J_{2}$ interactions%
\cite{chandra,nishi} and the third one ($\kappa $ term) is a three-spin
interaction.  
Note that this is the most general form 
which contains all possible short-range many-body spin-spin interactions. 
The interaction $V_{\rm{EQDM}}$ and the correspondence 
Fig.~\ref{fig:allowed-plaquette-states} determine 
the coefficients of the diagonal terms as\footnote{
$V_{\rm{EQDM}}$ is assigned per a plaquette.}:
\begin{eqnarray}
V_{\rm{EQDM}}^{\rm{(a)}} &= -4J+2J'+\lambda . \\
V_{\rm{EQDM}}^{\rm{(b)}} &= -J'. \\
V_{\rm{EQDM}}^{\rm{(c)}} &= J'. \\
V_{\rm{EQDM}}^{\rm{(d)}} &= -J.  \\
V_{\rm{EQDM}}^{\rm{(e)}} &= -2J+J'+\kappa .\
\label{eqn:V-terms}
\end{eqnarray}
Now it is clear why we have dropped in the previous section 
several dimer configurations allowed by the hardcore constraint alone 
(see Fig.~\ref{fig:allowed-plaquette-states}).  In fact, the pinwheel 
state $|0\rangle_1|0\rangle_2|0\rangle_3|0\rangle_4$ mentioned before 
yields zero when 
$V_{\rm{EQDM}}$ is applied.  
\totsuka{%
For the clarity of the argument, in what follows, we shall restrict 
our discussion to the diagonal parts of the Hamiltonians, although it is 
possible to write down the pseudo-spin interactions corresponding to 
the off-diagonal part of the EQDM. 
}

\begin{figure}[h]
\begin{center}
\includegraphics[width=12cm, clip]{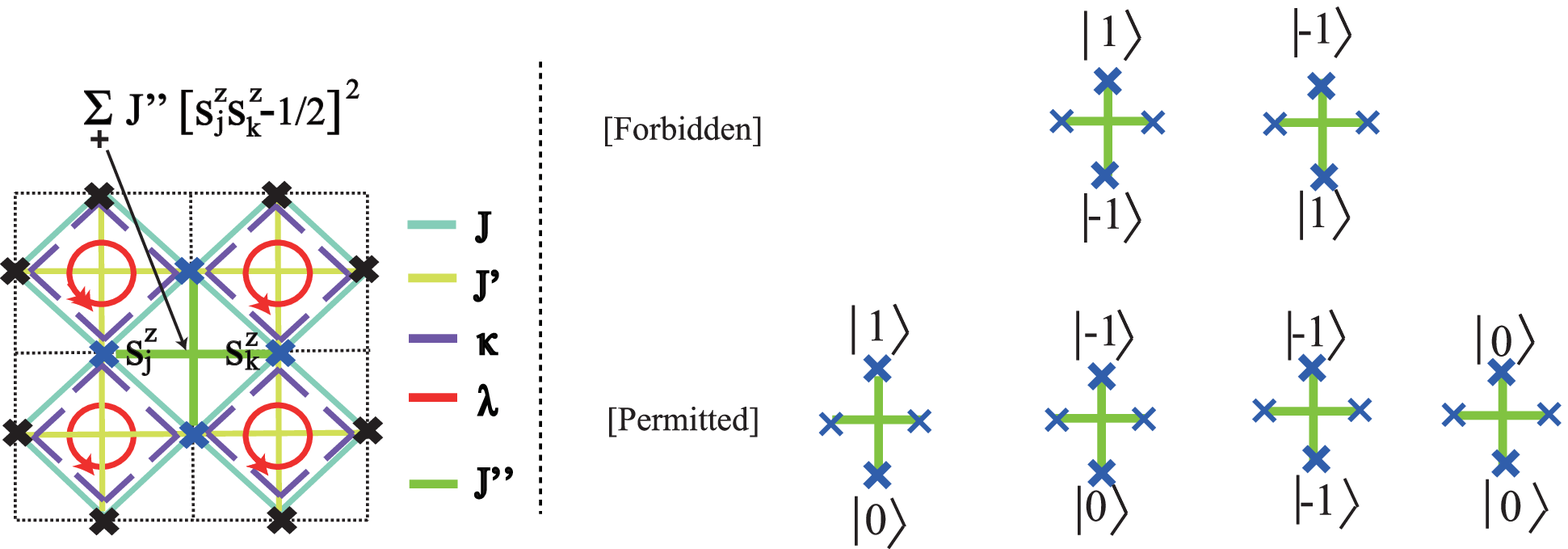}
\caption{(Color online) [Left] 
Interactions $V_{\rm EQDM}$ of the pseudo-spin  
Hamiltonian. \nakata{dotted dashed lines}
denote the original square lattice on which 
dimers are defined and the crosses marks the lattice sites where 
the pseudo-spins are located. 
The two-spin interactions $J$ and $J^{\prime}$ are similar 
to those of the $J_1$-$J_2$ model and $\lambda $ is the four-spin interaction 
necessary to realize the RK-point. $J^{\prime\prime}$ forces 
the one-to-one correspondence between the EQDM and the pseudo $S=1$
 Hamiltonian. 
[Right] Dimer configurations which are forbidden/permitted in
the EQDM, due to the definition of the three $S=1$ 
 eigenstates.\label{fig:forbidden-states}}
\end{center}
\end{figure}

The mapping is not completed yet at this stage;  
there exist configurations as shown in Fig.\ref{fig:forbidden-states} 
that are permitted by the constraint eq.(\ref{eqn:spin-cinstraint}) 
in the spin system, but not in the dimer systems 
because of the consistency of the definition Fig.\ref{fig:Def-S1-states} 
of the three eigenstates\footnote{%
The configuration where the two states $|1\rangle$ and $|-1\rangle$ 
are adjacent is not allowed since these are incompatible when on 
the neighboring bonds.}.
In order to suppress these {\em forbidden} states and 
realize one-to-one correspondence between the EQDM 
and the pseudo-spin Hamiltonian, we add the following 
interaction term to the spin Hamiltonian:
\begin{equation} 
V_{\rm{consistency}}=J^{\prime\prime} 
\sum_{+} \left[ s_j^z s_k^z  - \frac{1}{2} \right]^2     
\label{eqn:Vconst}    
\end{equation}
with $|J^{\prime\prime}| \gg | J |, | J^{\prime}|$, 
$J''>0$.
The summation is taken over all the vertices of the original lattice (see Fig.(\ref{fig:forbidden-states}).
The auxiliary interaction $V_{\rm{consistency}}$ excludes 
in the limit $J^{\prime\prime} \nearrow \infty$ 
the unwanted configurations ([Forbidden])
and selects the allowed ones ([Permitted]) while leaving the latter degenerate:
\begin{equation}
V^{\rm{Forbidden}} _{\rm{consistency}}=\frac{9}{4}J^{\prime\prime} \; , 
\quad
V^{\rm{Permitted}} _{\rm{consistency}}=\frac{1}{4}J^{\prime\prime} .\
\end{equation}
This finally establishes the one-to-one mapping  
between the diagonal part of the EQDM and that of the pseudo-spin Hamiltonian%
\footnote{%
It is evident that for a given dimer configuration we can determine 
the $S=1$ configuration {\em uniquely} by Fig.\ref{fig:Def-S1-states}. 
Given a pseudo-spin configuration, on the other hand, 
the constraint (\ref{eqn:spin-cinstraint}) and 
$V_{\rm{consistency}}$ guarantees that it satisfies the hardcore 
dimer constraint. Hence the mapping is one-to-one.}. 

\totsuka{%
The resulting pseudo spin Hamiltonian consists of the diagonal part and 
the auxiliary interaction $V_{\rm consistency}$ which imposes 
the consistency condition: 
}
\begin{eqnarray}
{V}_{\rm{pseudo}} &= & {V}_{\rm{EQDM}}+{V}_{\rm{consistency}}  \nonumber \\
&= & \sum_{\Box} [J s^z_i s^z_{i+1}+J's^z_i s^z_{i+2}
+\kappa s^z_i s^z_{i+1}s^z_{i+2} +\lambda s^z_i
 s^z_{i+2}s^z_{i+3}s^z_{i+4}] 
\nonumber \\%
& & + \sum_+ J''\left[s^z_j s^z_k-\frac{1}{2} \right]^2 
+ \rm{(rotations)} ,   
\label{eqn:pseudo-S-Ham}
\end{eqnarray}
where 
the parameters must satisfy $\mid J'' \mid $ $\gg $ $\mid J\mid $, 
$\mid J'\mid $.
By using the equivalence, the corresponding EQDM is then written down as:
\begin{eqnarray}
 {\cal H}_{\rm{EQDM}}^{{\rm{diagonal}}}= & \sum_{\Box}
\Biggl\{ 
 (-4J+2J^{\prime}+\lambda ) \,\configAhor\braAhor  -J \,\configD\braD  \nonumber\\
&-J^{\prime} \,\configB\braB  +(-2J+J'+\kappa ) \,\configE\braE \nonumber  \\
&+ J^{\prime} \,\configC\braC + (\rm{h.c.}) + (\rm{rotations}) 
+ (\rm{reflections})\Biggr\}.\
\end{eqnarray}

\nakata{As an effective Hamiltonian for Batista and Trugman's generalized $J_1$-$J_2$ Heisenberg model
(i.e. any states having at least one singlet dimer per plaquette are ground states),
the region of each parameters are restricted as } 
\begin{eqnarray}
V_{\rm{EQDM}}^{\rm{(d)}}<0\Leftrightarrow J>0.  \label{eqn:para1} \\
 V_{\rm{EQDM}}^{\rm{(e)}}<0\Leftrightarrow \kappa <2J-J'.\label{eqn:para2}\\
 V_{\rm{EQDM}}^{\rm{(a)}}+ V_{\rm{EQDM}}^{\rm{(b)}}<0, 
2V_{\rm{EQDM}}^{\rm{(a)}}+ V_{\rm{EQDM}}^{\rm{(c)}}>0. \\
\Leftrightarrow 4J-\frac{5}{2}J'<\lambda <4J-J', J'>0.
\label{eqn:para3}
\end{eqnarray}

Introducing the resonance terms 
\begin{eqnarray}
 {\cal H}_{\rm{EQDM}}^{\rm{resonance}}&= \sum_{\Box}
\Biggl\{ 
-t\, \configAvert \braAhor -t^{\prime}\,\configB\braBB 
-t^{\prime \prime}\,\configC\braCC  \nonumber \\
&\ \ \ \ \ \ \ \ \ \ \   + (\rm{rotations}) + (\rm{reflections})\Biggr\}.  
\end{eqnarray}
by hand, we arrive at the EQDM Hamiltonian (\ref{eqn:EQDM}).  
The coefficient $t$ describes the usual resonance of 
parallel dimers and the new term $t'$ ($t^{\prime\prime}$) moves 
three (four) dimers on a length-6 (8) loop. 
Thus, the EQDM is defined as 
${\cal H}_{\rm{EQDM}}={\cal H}_{\rm{EQDM}}^{\rm{resonance}}
+{\cal H}_{\rm{EQDM}}^{\rm{diagonal}}$. 
When $t^{\prime},t^{\prime \prime},J, J^{\prime}, \kappa=0$, 
the EQDM reduces to the usual Rokhsar-Kivelson QDM\cite{rk}.  
Furthermore, when $t=J^{\prime}=t^{\prime \prime}=\lambda =\kappa=0$, 
the phase transition from the columnar VBS to the herringbone, 
which has been studied by Papanikolaou et al.\cite{papa}, 
occurs.  

In summary, 
the requirement that the pseudo-spin Hamiltonian should consist only of 
short-range interactions enabled us to keep only a restricted class 
of dimer configurations out of infinitely many ones. 
This is the greatest advantage of using the pseudo-spin representation 
of EQDM.

\section{Calculation of $t'$ resonance effect in dimer smectic }
\label{sec:cal}

In this section, we outline the second-order calculation of the energy shift 
(eq.\ref{eqn:smectic}) caused by the $t'$ resonance effect for the dimer smectic.
Our calculation closely follows the work of Papanikolaou et al\cite{papa}. 

We treat the small positive resonance terms $t^{\prime}$ as perturbation:
\begin{eqnarray}
V_{\rm{perturb}}^{\rm{t'}}= & 
\sum_{\Box}\left\{-t^{\prime}\, \configB \braBB  + (\rm{rotations}) + (\rm{h.c.}) \right\} \; .
\end{eqnarray}
The first non-trivial contribution of $V_{\rm{perturb}}^{\rm{t'}}$ occurs 
at the second order and the energy shift due to the resonance is 
given by
\begin{eqnarray}
E_{\rm{resonance}}^{\rm{t'}}=\epsilon -{t'}^2 {\sum}^{\prime}_{m}
  \frac{V_{nm}V_{mn}}{\epsilon_m -\epsilon_n } +O({t'}^4),
\end{eqnarray}
\noindent
where $\epsilon$ is the unperturbed energy 
and the primed summation is over all dimer coverings except the original state. 
The terms in the sum which give nonzero contribution correspond to 
states connected to the initial state by a single flipped 
cluster; these processes are interpreted as brought about by 
quantum fluctuations.

\begin{figure}[h]
\begin{center}
\includegraphics[width=14cm, clip ]{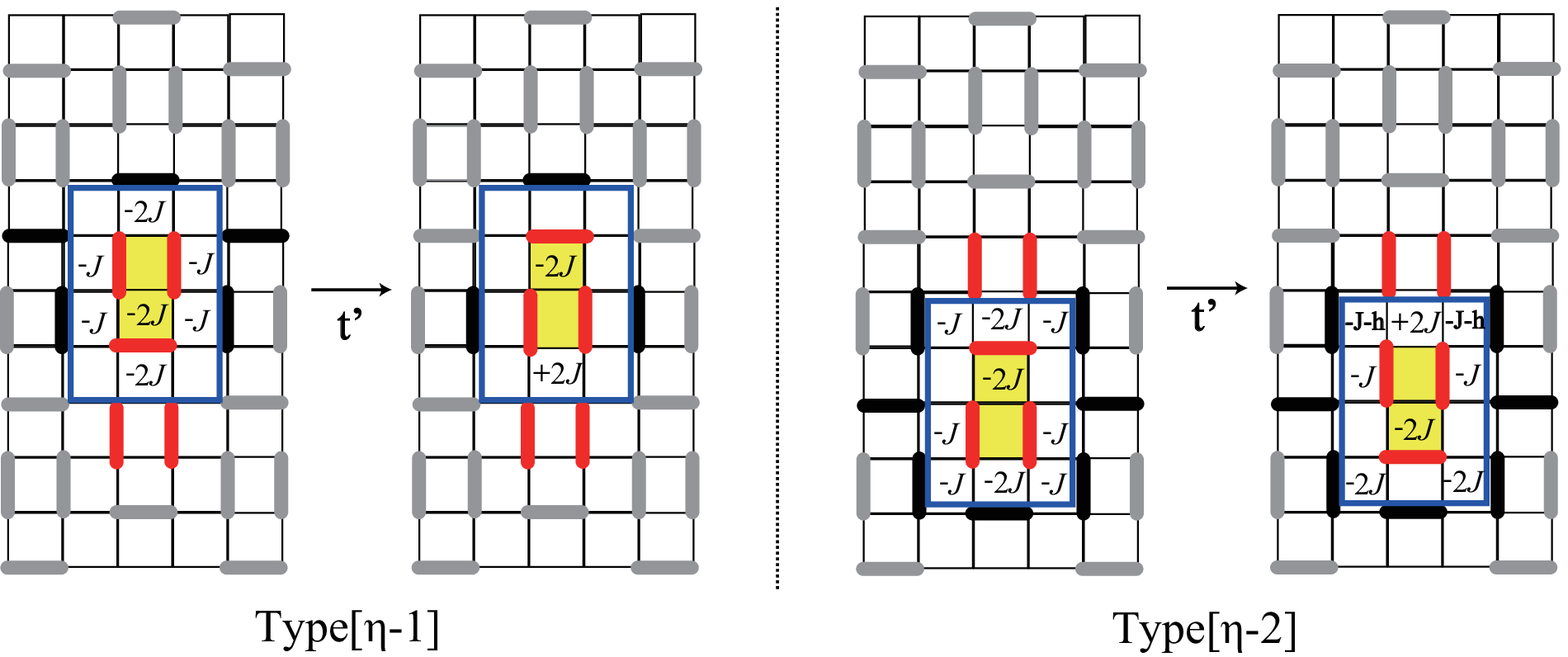}
\caption{(Color online) 
\totsuka{%
Typical intermediate states caused by $t'$ resonance term in type-($\eta$) dimer smectic.
There are two types of intermediate states (type[$\eta$-1] and [$\eta$-2]). 
The energies per plaquette are shown too.}
\label{fig:ita}}
\end{center}
\end{figure}
\begin{figure}[h]
\begin{center}
\includegraphics[width=8cm, clip ]{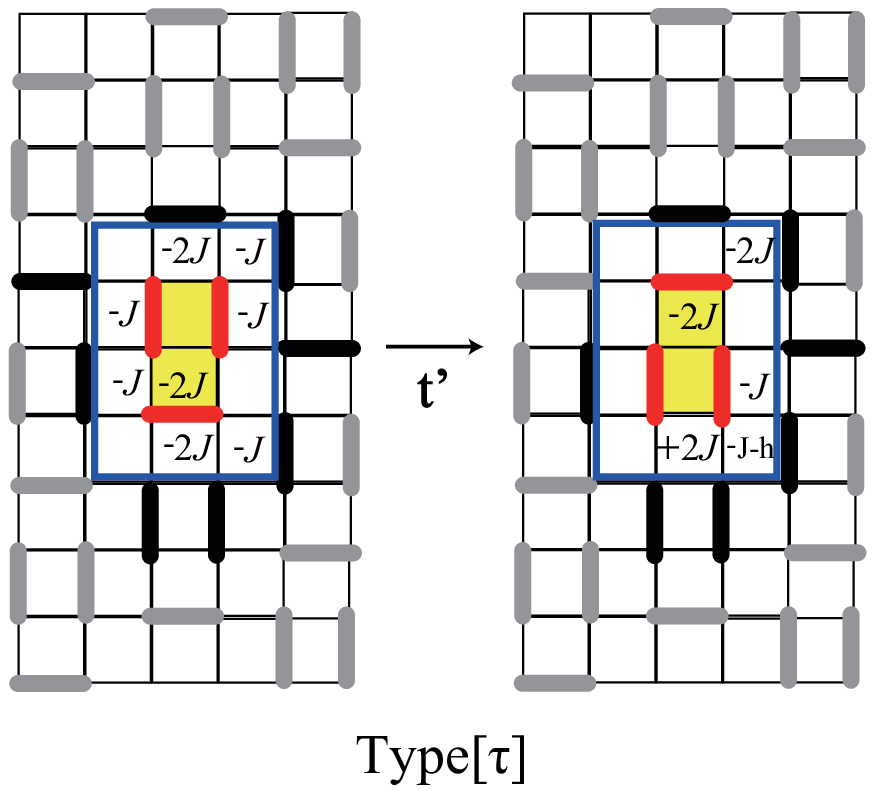}
\caption{(Color online) 
\totsuka{%
Typical intermediate states caused by $t'$ resonance term in type-($\tau$) dimer smectic.
}
\label{fig:tau}}
\end{center}
\end{figure}
Let us consider the energy shift for the dimer smectic case.
As has been discussed in section $\ref{sec:feature}$,
dimer smectic has two possible configurations of three neighboring 
columns (Fig.\ref{fig:two-config})
and therefore, we have to treat the two different patterns type-($\eta$) and ($\tau$) carefully.

\noindent%
\textit{Type}-($\eta $); 
First, we have to realize furthermore the two different ways 
(type[$\eta$-1] and type[$\eta$-2] in Fig.\ref{fig:ita}) 
how a cluster is flipped by $t'$.
In fact, the intermediate states are different and
their energies\footnote{$V_{\rm{EQDM}}$ (i.e. potential term)
is assigned per a plaquette as shown in Fig.\ref{fig:allowed-plaquette-states}.
}
are different from each other too;
\begin{eqnarray}
\Delta^{\rm{smectic}}_{\rm{t'}[\eta {\mathchar`-}1]} &=(0)-(-10J) =10J,
\end{eqnarray}
\begin{eqnarray}
\Delta^{\rm{smectic}}_{\rm{t'[\eta {\mathchar`-}2 ] }} &=(-8J-2h)-(-12J) =4J-2h.
\end{eqnarray}
\noindent%
\textit{Type-($\tau $)}; we do not have to worry about the above difference, 
because both cases give the same results. 
The energy shift is as below (Fig.\ref{fig:tau}):
\begin{eqnarray}
\Delta^{\rm{smectic}}_{\rm{t'[\tau ]}} &=(-4J-h)-(-11J) =7J-h.
\end{eqnarray}
Putting all these effects of the resonance $t^{\prime}$ together, 
we obtain the energy shift caused by $t'$-resonance of the dimer smectic as
\begin{eqnarray}
 \Delta E_{\rm{t'} \rm{resonance}}^{\rm{smectic}}&=
 - \frac{t'^2}{4J-2h}N_{\eta }\frac{N_y}{3} 
- \frac{t'^2}{10J}N_{\eta }\frac{N_y}{3}  
- \frac{t'^2}{7J-h}N_{\tau }\frac{2N_y}{3}.
\label{eqn:t'smectic}
\end{eqnarray}
The calculation goes similarly for $t$ and $t''$ as well
and collecting all the terms,
we obtain the energy  of the dimer smectic  with all resonance effects as eq.(\ref{eqn:smectic}).

\section*{References}
\bibliographystyle{unsrt}
\bibliography{JstatRef}

\end{document}